\documentclass[%
 reprint,
nofootinbib,
 amsmath,amssymb,
 aps,
]{revtex4-2}
\pdfoutput=1

\usepackage{graphicx}
\usepackage[usenames,dvipsnames]{xcolor}
\usepackage{dcolumn}
\usepackage{bm}
\usepackage{bbm}
\usepackage{hyperref}
\usepackage{float}

\usepackage{physics}

\usepackage{cleveref}
\DeclareMathOperator*{\argmin}{arg\,min}

\hypersetup{colorlinks=true}

\newcommand{\orcidauthor}[2]{\author{{\hypersetup{hidelinks}\href{https://orcid.org/#1}{#2}}}}

 \date{\today}

\begin{document}
\title{Correcting and extending Trotterized quantum many-body dynamics}

\orcidauthor{0000-0002-5891-3289}{Gian Gentinetta}
\email{gian.gentinetta@epfl.ch}
\orcidauthor{0000-0002-4745-7329}{Friederike Metz}
\orcidauthor{0000-0002-8887-4356}{Giuseppe Carleo}

\affiliation{%
Institute of Physics, \'Ecole Polytechnique F\'ed\'erale de Lausanne (EPFL), CH-1015 Lausanne, Switzerland \\
Center for Quantum Science and Engineering, EPFL, Lausanne, Switzerland
}%

\begin{abstract}
A complex but important challenge in understanding quantum mechanical phenomena is the simulation of quantum many-body dynamics. Although quantum computers offer significant potential to accelerate these simulations, their practical application is currently limited by noise and restricted scalability. In this work, we address these problems by proposing a hybrid ansatz combining the strengths of quantum and classical computational methods. Using Trotterization, we evolve an initial state on the quantum computer according to a simplified Hamiltonian, focusing on terms that are difficult to simulate classically. A classical model then corrects the simulation by including the terms omitted in the quantum circuit. While the classical ansatz is optimized during the time evolution, the quantum circuit has no variational parameters. Derivatives can thus be calculated purely classically, avoiding challenges arising in the optimization of parameterized quantum circuits.
We demonstrate three applications of this hybrid method. First, our approach allows us to avoid SWAP gates in the quantum circuit by restricting the quantum part of the ansatz to hardware-efficient terms of the Hamiltonian. Second, we can mitigate errors arising from the Trotterization of the time evolution unitary. Finally, we can extend the system size while keeping the number of qubits on the quantum device constant by including additional degrees of freedom in the classical ansatz.

\end{abstract}

\maketitle
\section{Introduction}

Solving the time-dependent Schrödinger equation for quantum many-body systems is an important piece in the puzzle of understanding nature at a microscopic scale~\cite{zhang2017,dborin2022,ebadi2021,altman2018,Jong2022,utility2023}. While analytical solutions quickly become unreachable as the number of particles increases, a plethora of computational methods are available to simulate the dynamics of such quantum systems up to a certain size. However, as the Hilbert space scales exponentially in the number of particles, classical simulation methods are limited to relatively small system sizes. Quantum computers comprise a promising tool for this application as the resources required for the simulation of quantum time evolution scale polynomially with the number of particles, in theory allowing for an advantage over classical algorithms~\cite{Feynman1982,ibm_17, chiesa_2019, google_2020, arute2020observation, neill2021}.

In practice, quantum simulations using Noisy Intermediate-Scale Quantum (NISQ) devices~\cite{Preskill2018quantumcomputingin} are restricted by limited scale and the effects of noise~\cite{childs2018,Babbush2018,Nam2019,Motta2021}. As a consequence, the simulation of quantum dynamics through Trotterization of the time evolution unitary~\cite{Hatano_2005,Berry_2006} is effectively only approximate: Due to the limited coherence time, only a small number of Trotter steps can be implemented on digital quantum computers, leading to significant Trotter errors. Further, the restricted connectivity of near-term quantum devices prohibits direct implementation of the time evolution of non-local Hamiltonians without additional SWAP gates further increasing the circuit depth. Finally, the small number of qubits available on current devices constrains the system sizes that can be simulated.

In recent years, variational quantum algorithms have been proposed to address the problem of limited coherence time~\cite{Bharti_2022}. By keeping the circuit depth constant, longer simulation times can be reached compared to a Trotterized time evolution at the cost of optimizing circuit parameters~\cite{ Yuan2019theoryofvariational,Benedetti2021, zhang2024adaptivevariationalquantumdynamics, Gacon2021,Barison_2021, linteau2023adaptiveprojectedvariationalquantum, Gentinetta2024overheadconstrained, gacon2023variational, gaconthesis}. The circuit ansatz can further be chosen to be hardware efficient, avoiding SWAP gates in the simulation of non-local Hamiltonians. The variational parameters are either updated using the time-dependent variational principle (TDVP)~\cite{dirac_1930, frenkel_1934, mclachlan_1964, YingTDVP, Yuan2019theoryofvariational,Benedetti2021, zhang2024adaptivevariationalquantumdynamics, Gacon2021} or by minimizing a loss function at every time step~\cite{Barison_2021, linteau2023adaptiveprojectedvariationalquantum, Gentinetta2024overheadconstrained,gacon2023variational}. In both cases, derivatives with respect to the variational parameters have to be computed on a quantum device, using e.g.~the parameter shift rule~\cite{Mitarai2018, Schuld_2019} which prohibits backpropagation scaling~\cite{abbas2023quantumbackpropagationinformationreuse}.

\begin{figure*}[t!]
    \centering
    \includegraphics[width=0.6\linewidth]{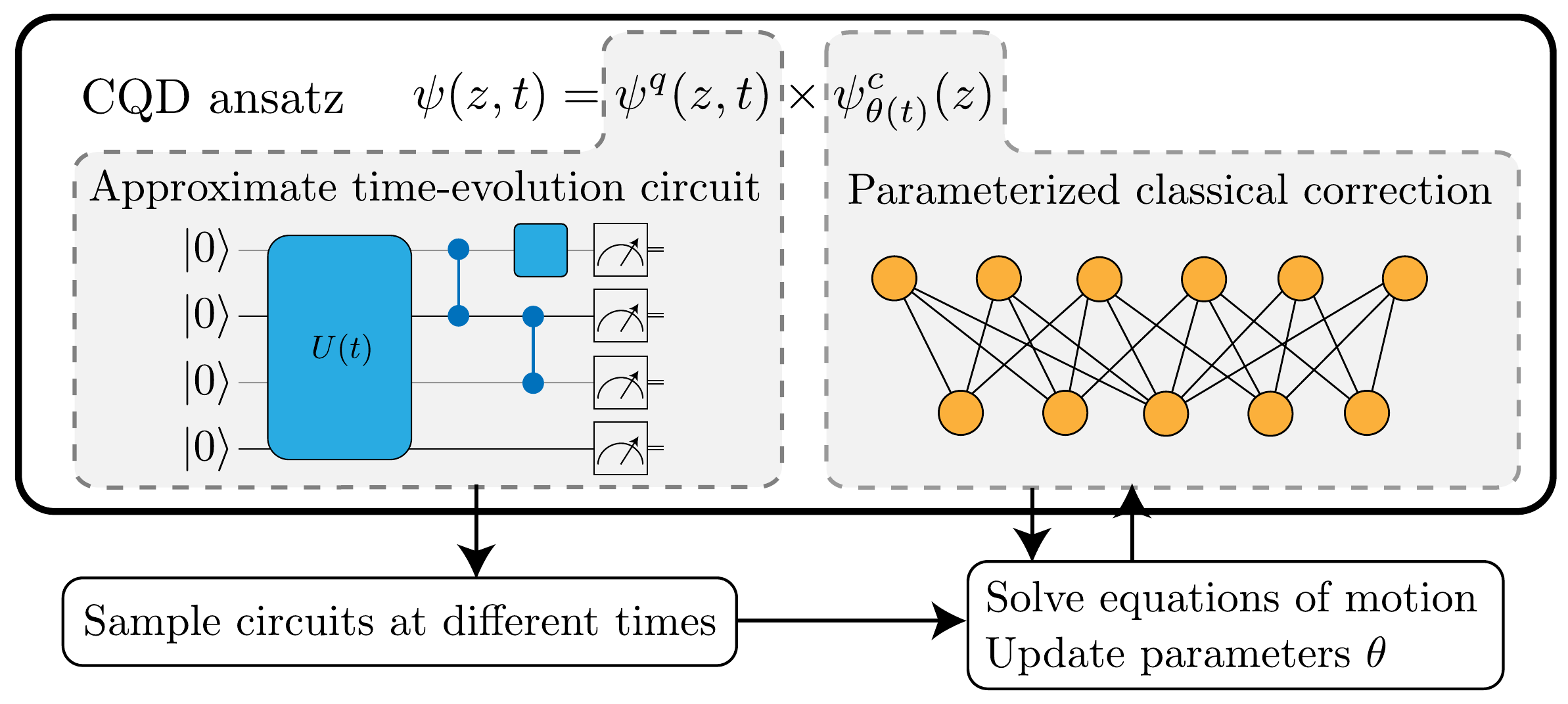}
    \caption{Sketch of the classically corrected quantum dynamics (CQD) ansatz. In order to evolve the ansatz in time according to a Hamiltonian $H$, a quantum circuit (on the left) is evolved with an approximate time evolution unitary $U(t)$. At every time step, the circuit is measured in multiple bases given by the Pauli decomposition of $H$ and the effective Hamiltonian $\tilde{H}$ underlying $U(t)$. Those samples, together with the classical ansatz (on the right) are used to obtain the equations of motion which are integrated in order to update the classical parameters $\theta$ iteratively.}
    \label{fig:ansatz}
\end{figure*}

On the other hand, to overcome the restrictions imposed by the limited number of available qubits on near-term devices, a range of quantum-classical hybrid algorithms has been proposed to split a large quantum system into smaller partitions~\cite{Bravyi_2016, Peng2020, Mitarai_2021,Mitarai2021overheadsimulating, piveteau2023circuit} for applications in groundstate problems~\cite{Fan2015, Gunst2017,yamazaki2018practical, Rossmannek_2021,Eddins_2022, Huembeli2022, paulin2023, barison2023embedding,XiaoHybridTensor2021,SunPerturbative2022, schuhmacher2024hybridtreetensornetworks} and more recently also for dynamics~\cite{gomez2023nearterm, Gentinetta2024overheadconstrained}. Crucially, the quantum circuits for the individual partitions remain parameterized and hence the challenges with computing derivatives remain. This is also the case for another class of hybrid algorithms, where a classical ansatz is combined with a parameterized quantum circuit to reach lower energies in discrete~\cite{Zhang22VarQNHybrid} or continuous~\cite{metz2024simulatingcontinuousspacesystemsquantumclassical} space groundstate problems.

In contrast, we propose a quantum-classical hybrid algorithm for the simulation of quantum dynamics that requires no optimization of parameters on the quantum circuit. Instead, we use Trotterized time evolution according to a simplified Hamiltonian containing terms that are difficult to simulate classically but can be efficiently implemented on the hardware. We then employ an optimizable classical ansatz to correct for the errors arising from the Trotterization as well as from the simplification of the Hamiltonian. As the quantum circuit is no longer parameterized, any derivatives can be computed purely classically using backpropagation. In fact, we show that all relevant quantities required for the optimization of the \textit{classically corrected quantum dynamics (CQD)} ansatz can be obtained by sampling the time-evolved quantum circuit in different bases and evaluating the classical part of the ansatz at those samples. 

The remaining article is structured as follows: In~\Cref{sec:methods}, we define the ansatz, show how expectation values are computed, and derive the equations of motion for the classical parameter updates. \Cref{sec:apps} showcases three applications of the method, demonstrating how the hybrid ansatz allows us to (i) correct Trotter errors [\Cref{sec:apps_trott}], (ii) avoid the use of SWAP gates by evolving with a hardware-efficient approximation of the Hamiltonian [\Cref{sec:app_hw_eff}], and (iii) extend the system size by treating additional degrees of freedom purely classically [\Cref{sec:apps_2subs}]. Finally, in~\Cref{sec:conclusion}, we discuss the results and provide an outlook of possible future applications of the method.

\section{Methods}
\label{sec:methods}

Our goal is to simulate the time evolution of an initial state $|\psi_0\rangle$ with respect to a time-independent Hamiltonian $H$. Ideally, we want to evolve the quantum state directly with $H$ on the quantum device, i.e., $\ket{\psi^q(t)} = e^{-iHt}\ket{\psi_0}$. However, consider the situation where the time evolution on the hardware can only be simulated with an approximate Hamiltonian $\tilde{H}$. This problem naturally occurs when, for example, the time evolution unitary is Trotterized. In these cases, we propose to use a classical parameterized function $\psi^c_{\theta(t)}(z)$ to correct the dynamics (see~\Cref{fig:ansatz}) and optimize it using an extended version of the time-dependent variational principle (TDVP)~\cite{YingTDVP}. Specifically, we sample the time-evolved quantum state at incremental time steps to obtain the appropriate equations of motion for the parameters $\theta(t)$ that are then updated accordingly.

\subsection{Hybrid ansatz\label{sec:hybrid_ansatz}}
We define the Classically Corrected Quantum Dynamics (CQD) ansatz at time $t$ in the computational basis spanned by bitstrings $z$ as
\begin{equation}
    \label{eq:ansatz}
    \psi_{\theta(t)}\left(t,z\right) = \bra{z}\ket{\psi^q(t)}\psi_{\theta(t)}^c(z) = \psi^q(t,z)\psi_{\theta(t)}^c(z),
\end{equation}
where $|\psi^q(t)\rangle = U(t)\ket{\psi_0}$ is a quantum state implemented as a quantum circuit, with $U(t)$ a time evolution unitary specified later, and $\psi^c$ is the classical correction such as a mean-field, Jastrow, or a neural network ansatz. Importantly, the classical contribution to the wave function is parameterized with time-dependent parameters $\theta(t)$, while the time dependence of the quantum contribution is explicitly given through a deterministic time evolution of the initial state. Thus, the quantum circuit contains no optimizable parameters. Note that, in general, the hybrid wave function $\psi_{\theta(t)}\left(t,z\right)$ is not required to be normalized.

We now show how expectation values of observables can be computed within the general CQD ansatz of~\eqref{eq:ansatz} while the time evolution algorithm and parameter optimization will be discussed in the next section. As any observable can be decomposed into a sum of Pauli strings $P \in \{\mathbb{I},X,Y,Z\}^{\otimes n}$, it suffices to consider a single arbitrary Pauli string $P$ whose expectation value yields
\begin{equation}
    \label{eq:pauli}
    \bra{\psi_\theta}P\ket{\psi_\theta} = \sum_{z,z'}\psi^q(t,z)^*\psi^c_\theta(z)^*\psi^q(t,z')\psi^c_\theta(z')\bra{z}{P}\ket{z'}. 
\end{equation}
For diagonal Pauli strings $P\ket{z} = S(z)\ket{z}$, this can be immediately written as
\begin{align}\begin{split}
\label{eq:diagonal_pauli}
\bra{\psi_\theta}P\ket{\psi_\theta} &= \sum_z  |\bra{z}\ket{\psi^q}|^2S(z)|\psi^c_\theta(z)|^2 \\
&= \mathbb{E}_{z \sim |\bra{z}\ket{\psi^q}|^2}\left[S(z)|\psi^c_\theta(z)|^2\right] \\
&\approx \frac{1}{N_s}\sum_{z \sim |\bra{z}\ket{\psi^q}|^2}S(z)|\psi^c_\theta(z)|^2,
\end{split}\end{align} 
where we dropped the time dependence for readability. The expectation value can hence be estimated by evaluating $S(z)|\psi^c_\theta(z)|^2$ classically at $N_s$ configurations $z$ sampled from the quantum circuit $|\psi^q\rangle$.

For non-diagonal observables, Zhang et al.~\cite{Zhang22VarQNHybrid} show how a more intricate measurement scheme can be exploited to similarly sample bitstrings $z$ from a quantum circuit and evaluate the classical correction accordingly. In \Cref{app:expvals}, we discuss the sampling procedure in detail. We show that for any Pauli string $P$ and classical function $f$, expressions of the form
\begin{equation}
\label{eq:pauli_expect}
    \sum_{z,z'} \psi^q(t,z)^*\psi^q(t,z')\bra{z}{P}\ket{z'}f(t,z, z')
    \end{equation}
can be computed by sampling configurations $z$ form $|\psi^q\rangle$ in at most 2 bases. Evaluating the expectation value of a Pauli string for the CQD ansatz~\eqref{eq:pauli} is a special case of this result with $f(t,z,z') = (\psi^c_{\theta(t)}(z))^*\psi^c_{\theta(t)}(z')$ in~\eqref{eq:pauli_expect}. 

\subsection{Derivation of the equations of motion}
Our goal is to optimize the CQD ansatz at incremental time steps such that the resulting hybrid wavefunction follows the desired time evolution according to a Hamiltonian $H$. For an infinitesimal time step $\delta t$ we define
\begin{equation}
    \psi_{\theta(t + \delta t)}(t + \delta t, z) = \psi^c_{\theta(t) + \delta \theta}(z)\bra{z}e^{-i\delta t \tilde{H}}\ket{\psi^q(t)},
\end{equation}
where $\delta\theta$ is the update to the classical parameters that is to be determined. To that end, we apply McLachlan's variational principle~\cite{mclachlan_1964,YingTDVP}
\begin{equation}
    \label{eq:mclachlan}
    \delta \theta^* = \argmin_{\delta\theta} \left\Vert e^{-i\delta t H}\ket{\psi_\theta\left(t\right)} - \ket{\psi_{ \theta + \delta \theta}\left(t +\delta t\right)}  \right\Vert^2.
\end{equation}
The main steps to solve this minimization problem are outlined below assuming that the ansatz is normalized ($\braket{\psi} = 1$). The results are later generalized to the case of unnormalized wave functions.

Using the shorthand notation $\partial_k = \frac{\partial}{\partial_{\theta_k}}$, we note that the derivative with respect to time can be expanded as
\begin{equation}
    \label{eq:dt}
    \frac{d}{dt} = \partial_t +\sum_k  \frac{\partial\theta_k}{\partial_t}\partial_k= \partial_t +\sum_k  \dot{\theta}_k\partial_k.
\end{equation}
Next, we Taylor expand both the time evolution unitary 
\begin{equation}
    e^{-i\delta t H} = \mathbbm{1} - i\delta t H + O(\delta t^2)
\end{equation}
and the ansatz
\begin{equation}
    \ket{\psi_{ \theta + \delta \theta}\left(t +\delta t\right)} = (1 + \delta t \frac{d}{dt})\ket{\psi_\theta\left(t\right)} + O(\delta t ^2) .
\end{equation}
Inserting these expressions into~\eqref{eq:mclachlan} and using $\delta \theta_k = \dot{\theta}_k\delta t$, we obtain
\begin{widetext}
\begin{align}\begin{split}
    \label{eq:tdvp_deriv}
    \delta \theta^* &= \delta t\argmin_{\dot{\theta}} \left\Vert \left(\mathbbm{1} - i\delta t H\right)\ket{\psi} - \left(1 + \delta t \partial_t +\delta t\sum_k  \dot{\theta}_k\partial_k\right)\ket{\psi}  \right\Vert^2 \\
    &= \delta t\argmin_{\dot{\theta}}\left\Vert i H\ket{\psi} +  \sum_k\dot{\theta_k} \ket{\partial_k\psi}  + \ket{\partial_t\psi} \right\Vert^2\\
    &=\delta t\argmin_{\dot{\theta}} \sum_k\dot{\theta}_k\left[i\bra{\partial_k\psi}H\ket{\psi} + \bra{\partial_k\psi}\ket{\partial_t\psi} + \sum_{l} \dot{\theta}_l(\bra{\partial_l\psi}\ket{\partial_k\psi} + c.c.\right],
\end{split}\end{align}
\end{widetext}
where we use the shorthand $\ket{\psi} = \ket{\psi_\theta(t)}$ and all terms not depending on $\dot{\theta}$ have been dropped as they do not contribute to the $\argmin$.

To perform the minimization, we set the derivative of the objective function with respect to $\dot{\theta}$ to zero and obtain the equations of motion for the parameter changes $\dot{\theta}$.
\begin{equation}
    \label{eq:eom}
    \sum_l\!\dot{\theta}_l \Re{\bra{\partial_k\psi}\ket{\partial_l\psi}} \!= \Im{\bra{\partial_k\psi}H\ket{\psi}} -\Re{\bra{\partial_k\psi}\ket{\partial_t\psi}}.
\end{equation}
The generalized equations of motion for an unnormalized ansatz are given by (see~\Cref{app:unnormalized} for the derivation)
\begin{equation}
    \label{eq:eom_norm}
    \sum_{l}S_{kl}\dot{\theta}_l = C_k - T_k,
\end{equation}
where 
\begin{equation}
\label{eq:unnorm_s}
    S_{kl} = \Re{\frac{\bra{\partial_k\psi}\ket{\partial_k\psi}}{\braket{\psi}} - \frac{\bra{\partial_k\psi}\ket{\psi}\bra{\psi}\ket{\partial_l\psi}}{\braket{\psi}^2}}
\end{equation}
is the quantum geometric tensor (QGT),
\begin{equation}
\label{eq:unnorm_c}
C_k = \Im{\frac{\bra{\partial_k\psi}H\ket{\psi}}{\braket{\psi}} - \frac{\bra{\partial_k\psi}\ket{\psi}\bra{\psi}H\ket{\psi}}{\braket{\psi}^2}}
\end{equation}
are the forces usually appearing in the equations of TDVP~\cite{mclachlan_1964} and
\begin{equation}
\label{eq:unnorm_t}
T_k = \Re{\frac{\bra{\partial_k\psi}\ket{\partial_t\psi}}{\braket{\psi}} - \frac{\bra{\partial_k\psi}\ket{\psi}\bra{\psi}\ket{\partial_t\psi}}{\braket{\psi}^2}}
\end{equation}
are additional forces arising from the explicit time dependence of the ansatz. 

The equations of motion~\eqref{eq:eom_norm} can be solved using any numerical ODE integrator, such as the Euler method, to compute the parameter updates $\theta(t+\delta t)$. 
The TDVP time step size $\delta t$ and the choice of integrator are hyperparameters of the optimization and must be selected to balance computational cost and simulation stability.

\subsection{Calculating the forces and the quantum geometric tensor}
\label{sec:calc_forc_and_qgt}
In~\Cref{app:der_table} we show that all quantities appearing in the equations of motion~\eqref{eq:unnorm_s}-\eqref{eq:unnorm_t} can be written in the form of~\eqref{eq:pauli_expect}. Each term can thus be computed by sampling configurations from quantum circuits (potentially in different bases) and evaluating classical functions on said samples.

As an example, the term containing the explicit time derivative appearing in the forces $T_k$ is calculated as
\begin{align}\begin{split}
    \braket{\psi}{\partial_t\psi}\!\! &=\! \sum_z \psi^q(t,z)^*(\partial_t \psi^q(t,z))\psi^c_\theta(z)^*\psi^c_\theta(z)\\
    &=\! \sum_z \psi^q(t,z)^*\bra{z}(-i\tilde{H})\ket{\psi^q(t)}|\psi^c_\theta(z)|^2 \\
    &=\! -i\sum_{z,z'}\psi^q(t,z)^*\bra{z}\tilde{H}\ket{z'}\psi^q(t,z')|\psi^c_\theta(z)|^2 \\
    &=\! -i\!\sum_k \!\! w_k\!\sum_{z,z'}\!\!\psi^q(t,z)^*\psi^q(t,z')\!\bra{z}{P}_k\ket{z'}\!|\psi^c_\theta(z)|^2,
\end{split}\end{align}
where we decomposed $\tilde{H}=\sum_kw_k{P}_k$. This fits into the form of~\eqref{eq:pauli_expect} with $f(t,z,z') = |\psi^c_\theta(z)|^2$. The corresponding results for all other terms appearing in the expressions of the QGT and forces are summarized in~\Cref{tab:pauli}.

At every time step the circuit $\ket{\psi^q(t)}$ has to be sampled in a number of bases
\begin{equation}
    N_{\text{bases}} = 2N_{nd}(H + \tilde{H}) + 1,
\end{equation}
where $N_{nd}(H + \tilde{H})$ are the number of non-diagonal Pauli strings in the Pauli decomposition of $H$ and $\tilde{H}$, respectively. The factor of 2 appears as different bases have to be chosen to obtain the real and imaginary part of the expectation values. Note that if Pauli strings are contained in both Hamiltonians, the samples can be reused and hence, $N_{nd}(H + \tilde{H}) \leq N_{nd}(H) + N_{nd}(\tilde{H})$.

\begin{table}[h]
    \centering
    \renewcommand{\arraystretch}{1.3}
    \begin{tabular}{|c|c|c|}
        \hline
        \textbf{Expression} & \textbf{Pauli Strings} & \boldmath$f(t, z, z')$ \\
        \hline
        $\braket{\psi}$ & Identity & $|\psi^c_\theta(z)|^2$ \\
        \hline
        $\bra{\partial_l\psi}\ket{\partial_k\psi}$ & Identity & $(\partial_l\psi^c_\theta(z))^*\partial_k\psi^c_\theta(z)$ \\
        \hline
        $\bra{\partial_k\psi}\ket{\psi}$ & Identity & $(\partial_k\psi^c_\theta(z))^*\psi^c_\theta(z)$ \\
        \hline
        $\bra{\psi}H\ket{\psi}$ & $P \in \mathcal{P}(H)$  & $(\psi^c_\theta(z))^*\psi^c_\theta(z')$ \\
        \hline
        $\bra{\partial_k\psi}H\ket{\psi}$ & $P \in \mathcal{P}(H)$  & $(\partial_k\psi^c_\theta(z))^*\psi^c_\theta(z')$ \\
        \hline
        $\braket{\psi}{\partial_t\psi}$ & $P \in \mathcal{P}(\tilde{H})$  & $|\psi^c_\theta(z)|^2$ \\
        \hline
        $\braket{\partial_k\psi}{\partial_t\psi}$ & $P \in \mathcal{P}(\tilde{H})$  & $(\partial_k\psi^c_\theta(z))^*\psi^c_\theta(z)$ \\
        \hline
    \end{tabular}
\caption{All the quantities appearing in the equations of motion can be recast to expressions of the form~\eqref{eq:pauli_expect}. Here, we list the Pauli strings according to which the time evolved circuit has to be sampled and the classical functions that have to be evaluated on those samples to compute each expression. We write $\mathcal{P}(H)$ to denote the set of Pauli strings appearing in the decomposition of $H$.}
    \label{tab:pauli}
\end{table}

\section{Applications}
\label{sec:apps}
In this section, we present three use cases that illustrate how the CQD ansatz can be leveraged to correct or extend Trotterized quantum dynamics. First, we demonstrate its ability to mitigate Trotter errors. Second, we show how the CQD framework enables hardware-efficient time evolution on a quantum computer, even for Hamiltonians with non-local terms. Finally, we explore how additional degrees of freedom can be incorporated entirely within the classical component of the ansatz, eliminating the need for extra qubits on the quantum device. Note that these examples are not exhaustive but highlight the versatility and potential of the CQD ansatz for quantum dynamics simulations.

\subsection{Correcting Trotter errors}
\label{sec:apps_trott}
In order to implement the time evolution unitary $U(t) = \exp(-iHt)$ for general Hamiltonians as a quantum circuit, we usually employ a Trotterization of $U(t)$. For a Hamiltonian with non-commuting terms $[H_k, H_l] \neq 0$ we can expand for example to first order
\begin{equation}
    \label{eq:Trotter}
    \exp(-iHt) = \left(\prod_k \exp\left(-iH_k\Delta t\right)\right)^{t/\Delta t} + O(t^2\Delta t),
\end{equation}
which is a good approximation for small Trotter steps $\Delta t$. The benefit of this Trotterization is that individual terms can now be efficiently implemented as a quantum circuit through a product of locally acting quantum gates. In practice, on noisy devices we only have access to a limited number of Trotter steps due to the restricted coherence time. This leads to a trade-off between Trotter error and hardware noise. 

\begin{figure}[t!]
    \centering
    \includegraphics[width=\linewidth]{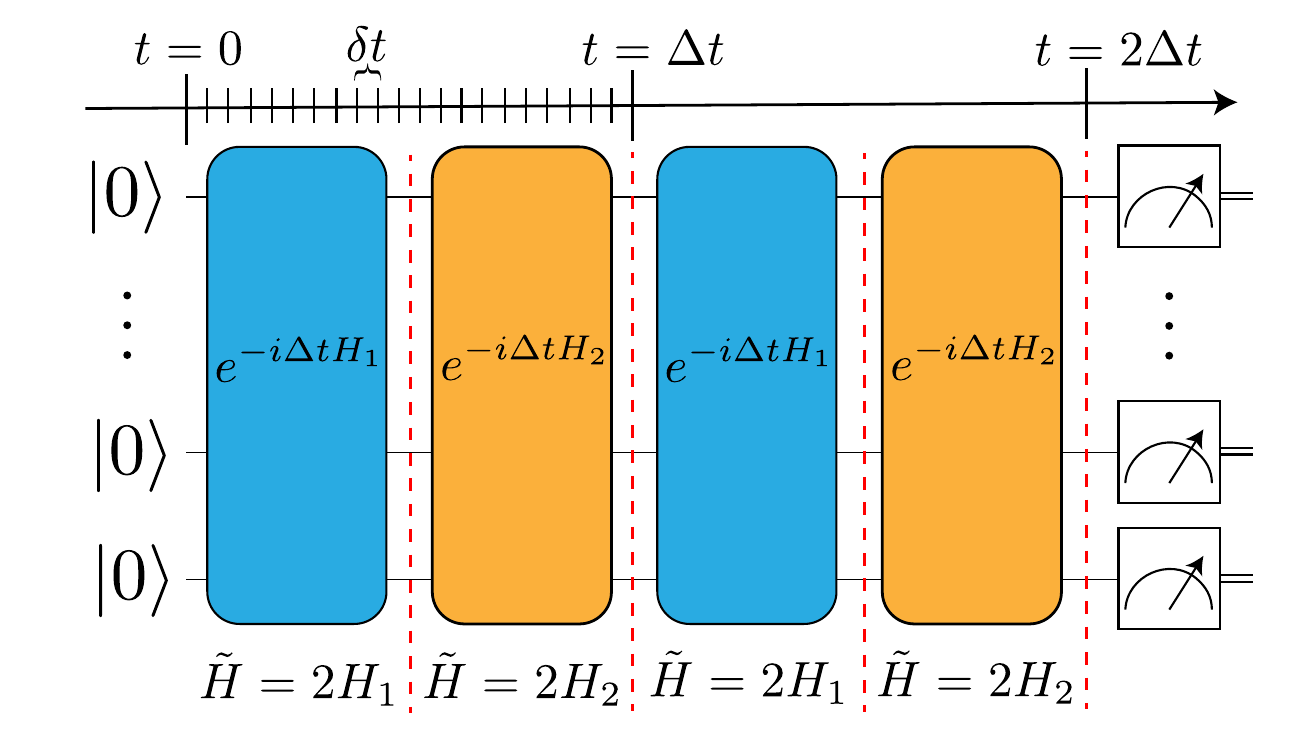}
    \caption{Quantum circuit representing two first-order Trotter steps of a Hamiltonian $H = H_1 + H_2$, where $H_1$ and $H_2$ are two non-commuting terms. To correct for the error arising from the large Trotter steps $\Delta t$, we update the classical ansatz at a smaller timescale $\delta t$. Within this timescale, the unitary $e^{-i\Delta tH_i}$ can be seen as a time evolution according to an effective Hamiltonian $\tilde{H} = 2H_i$ for a time $\Delta t / 2$.}
    \label{fig:Trotter_sketch}
\end{figure}

Here, we demonstrate how our hybrid method can be used to address this problem. To this end, we can interpret Trotterized time evolution as exact quantum evolution with an effective piece-wise constant Hamiltonian (see~\Cref{fig:Trotter_sketch}). For example, the first order Trotter decomposition for a Hamiltonian $H = H_1 + H_2$ is written as
\begin{align}\begin{split}
    \ket{\psi(t+\Delta t)}&=\exp(-iH\Delta t)\ket{\psi(t)}\\
    &\approx \exp(-iH_2\Delta t)\exp(-iH_1\Delta t)\ket{\psi(t)} \\
    &=\exp(-2iH_2\frac{\Delta t}{2})\exp(-2iH_1\frac{\Delta t}{2})\ket{\psi(t)}.
\end{split}\end{align}
A Trotter step can thus be described within our framework by setting $\tilde{H} = 2H_1$ for the first half of the time step $\Delta t$ and $\tilde{H} = 2H_1$ during the second half. This generalizes for multiple Trotter steps, Hamiltonians with more than two terms, and higher order Trotter decompositions. Given an expressive enough classical ansatz $\psi^c_\theta$ and a TDVP timescale $\delta t \ll \Delta t$, we can thus correct errors arising from coarse Trotterization. 

As an example, we simulate the transverse-field Ising model (TFIM) on a chain of $L=10$ spins with periodic boundary conditions, described by the Hamiltonian
\begin{equation}
    \label{eq:tfim}
    H = -h\sum_{i=1}^LX_i - J\sum_{i=1}^{L}Z_{i}Z_{i+1},
\end{equation}
where we set the local field $h=1$ and the coupling strength $J=2$. $X_i$ and $Z_i$ denote the corresponding Pauli operators acting on the $i$-th qubit. We start in the +1 eigenstate of the $X$ basis $\ket{\psi_0} = \ket{+}^{\otimes L}$ and evolve it in time using Trotterization to second order with a Trotter step size $\Delta t = 0.25$. The TDVP time step is set to $\delta t = 0.005$ and we integrate the equations of motion using the Euler method. Note that here we can in fact split the Hamiltonian into two non-commuting terms $H = H_x + H_z$, where $H_x$ contains the local fields whereas $H_z$ consists of the couplings. 

As the classical model we choose a simple Jastrow ansatz~\cite{jastrow1955} given by
\begin{equation}
    \label{eq:jastrow}
    \psi^c_\theta(z) = \exp\left(\sum_i\theta_iz_i + \sum_{ij}\theta_{ij}z_iz_j\right),
\end{equation}
where all parameters are complex and the two-body parameters are symmetric $\theta_{ij} = \theta_{ji}$. For $L$ spins, we count a total of $L + \frac{L(L-1)}{2}$ complex parameters. The parameters are initialized to $\theta(0) = 0$, such that $\psi^c_{\theta(0)}(z) \equiv 1$.

\begin{figure*}[t!]
    \centering
    \includegraphics[width=\linewidth]{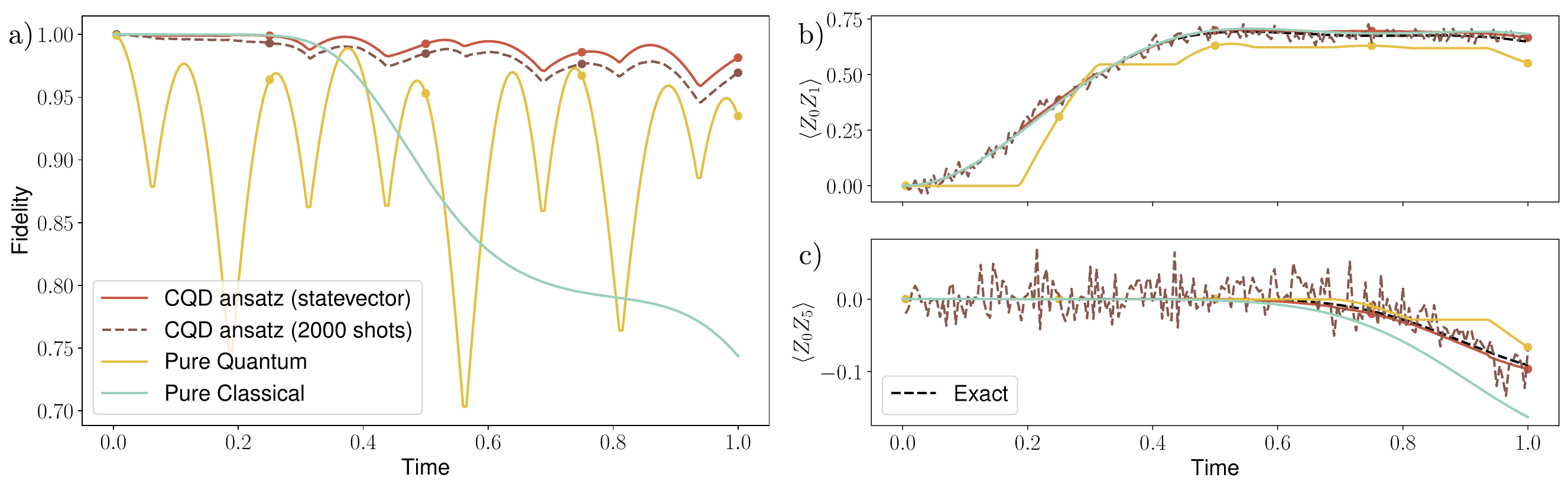}
    \caption{\textbf{Correcting Trotter errors} in the simulation of the TFIM on a chain of 10 spins with periodic boundary conditions. (a) The fidelity of the simulated quantum states with respect to the exact solution. We compare a purely classical simulation based on a Jastrow ansatz (blue line) as defined in~\Cref{eq:jastrow} with the CQD ansatz combining the Jastrow with the Trotterized quantum circuit for both a noiseless (red line) and a shot based simulation (brown dashed line). The uncorrected quantum simulation is displayed for reference (yellow line). (b) $ZZ$ correlation between neighboring qubits and (c) two qubits furthers apart. The exact expectation values are shown in black. The Trotterization is chosen at second order with time step size $\Delta t = 0.25$ (full Trotter steps are indicated with dots).}
    \label{fig:correct_Trotter_fid}
\end{figure*}

In~\Cref{fig:correct_Trotter_fid}~(a), we plot and compare the fidelities of quantum states simulated by different means with respect to the exact solution
\begin{equation}
    \label{eq:fid}
    F(\ket{\psi_{\theta}(t)},\ket{\psi_{exact}}) = \left|\bra{\psi_{\theta}(t)}e^{-iHt}\ket{\psi_0}\right|^2.
\end{equation}

Simulating the evolution using solely the Trotterized quantum circuit approximation without the classical model (yellow line) gives rise to oscillating, non-smooth fidelities. Cusps appear whenever the piecewise constant evolution is switched between $H_x$ and $H_z$. The points in time where a full Trotter step is completed are marked with dots. In contrast, simulating the system purely with a classical Jastrow ansatz without the help of the quantum circuit (blue line) results in a smooth fidelity curve, which however quickly drops for times $t\gtrsim0.3$. Hence, the Jastrow ansatz alone is not expressive enough to capture the dynamics of the system at large times.

Finally, we combine the Trotterized circuit and the Jastrow ansatz within our CQD framework. We show both an ideal statevector simulation (red solid line) and a shot based simulation (brown dashed line) with 1000 shots per evaluated circuit\footnote{The simulation is performed using shot noise. However, to compute the fidelities we use the full statevectors.}. In each case we obtain fidelities superior to the purely quantum and purely classical simulation for $t \gtrsim 0.3$. Hence, the CQD ansatz allows for an accurate simulation using a classical ansatz with fewer parameters that would otherwise not be sufficient to capture the full dynamics of the system alone. Furthermore, the non-smooth, oscillating behavior stemming from the Trotterization is washed-out resulting in high fidelities of the simulated CQD state also at intermediate times in-between Trotter steps.

To obtain a better understanding of the expressibility of the ansätze, we additionally plot two observables. \Cref{fig:j1j2_fid}~(b) shows the correlator $\langle Z_0Z_1\rangle$ of two neighboring spins. For this local observable, the Jastrow ansatz alone is able to capture the dynamics faithfully for the entire evolution. However, when we examine the correlator $\langle Z_0Z_5\rangle$ between two spins furthest apart, we observe that only the CQD ansatz provides an accurate prediction over the entire time window (see \Cref{fig:j1j2_fid}~(c)). This indicates that the quantum circuit is required for the simulation of long-range correlations whereas the Jastrow ansatz is needed to correct the arising Trotter error and interpolate the dynamics between the discrete Trotter steps. The latter is apparent for both observables; the Trotterized evolution without the classical correction gives rise to piecewise constant expectation values whenever the state is evolved according to the Hamiltonian $H_z$ (as this Hamiltonian commutes with the observable). However, even at times that match a full Trotter step (yellow dots), the simulated expectation values are less accurate than those predicted by the CQD ansatz due to the Trotter error. Note that the statistical noise apparent in the shot based simulation could be alleviated by increasing the number of shots.

\subsection{Hardware efficient time evolution}
\label{sec:app_hw_eff}
Qubit connectivity in near-term quantum hardware, such as for super-conducting qubits, is often restricted by the topology given by the device. For example, IBM Quantum hardware follows a heavy-hex topology~\cite{ibm_roadmap} and Google's willow processor is connected in a square lattice~\cite{willow2024}. Applying two-qubit gates between qubits that are not connected with native hardware connections requires additional SWAP gates which greatly increases the circuit depth and thus the required coherence time to successfully run the circuit. This problem arises for example when evolving a quantum state in time with respect to a Hamiltonian that does not follow the hardware topology. Here, we propose to evolve the quantum state with a hardware-efficient approximation of the Hamiltonian. We only include those terms of $H$ that respect the topology of the device in $\tilde{H}$, where $\tilde{H}$ again denotes the effective Hamiltonian that governs the dynamics on the quantum computer. We then use a classical ansatz to correct for the non-local terms in $H$ that have been omitted in the approximation $\tilde{H}$.
\begin{figure}[t!]
    \centering
    \includegraphics[width=\linewidth]{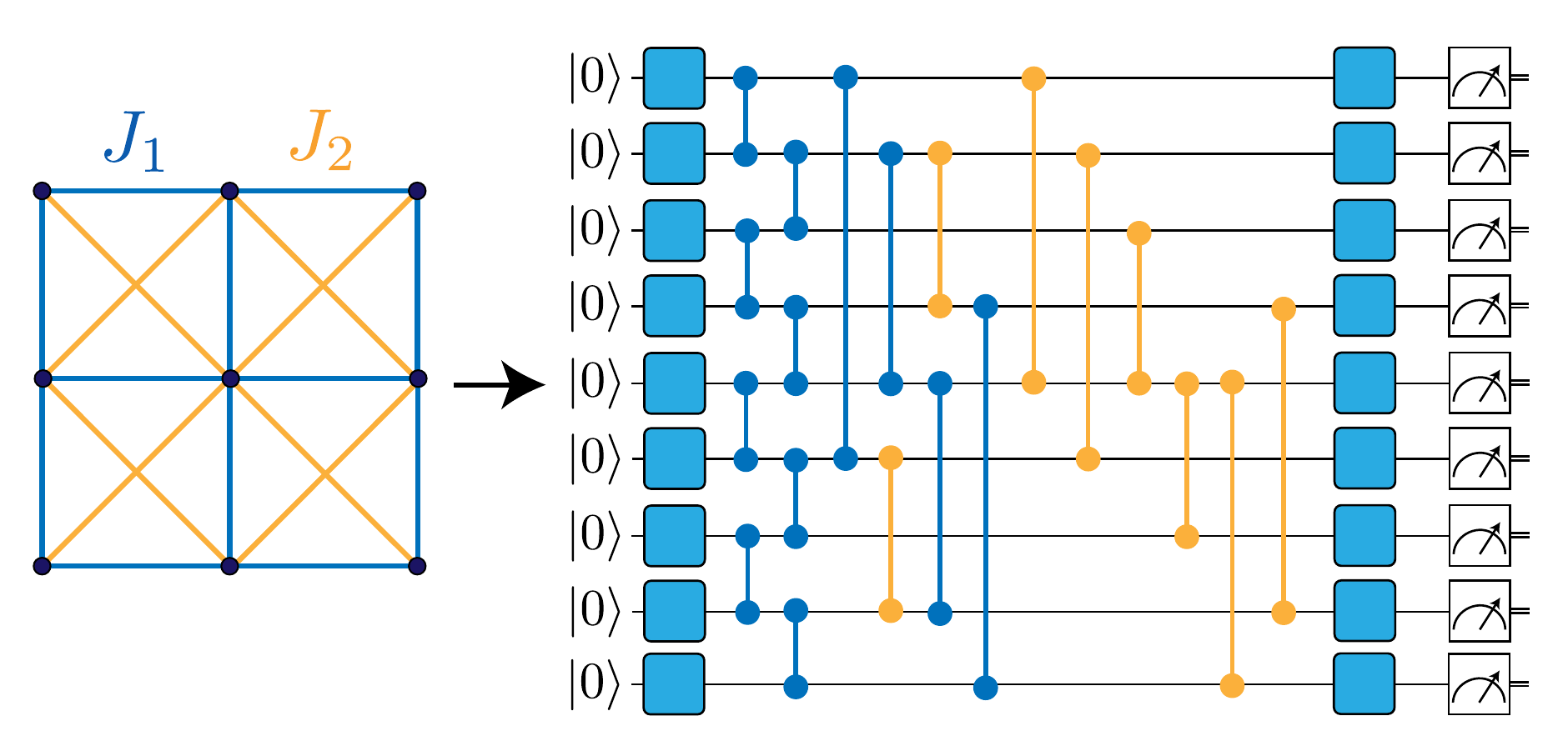}
    \caption{Two-dimensional spin system with strong nearest neighbor ($J_1$) and weak next-nearest neighbor interaction ($J_2$). The left panel shows the topology of the system where the nodes indicate individual spins with a local field $h$ and the edges represent the couplings $J_{ij} Z_i\otimes Z_j$. The right panel shows the circuit implementing a single second-order Trotter step for such a system with boxes representing the $R_X$ rotations and lines between qubits the $R_{ZZ}$ gates. Even without including the required SWAP gates, the circuit depth can be significantly reduced when omitting the next-nearest neighbor couplings (yellow lines/gates).}
    \label{fig:j1j2_explain}
\end{figure}
\begin{figure*}[t!]
    \centering
    \includegraphics[width=\linewidth]{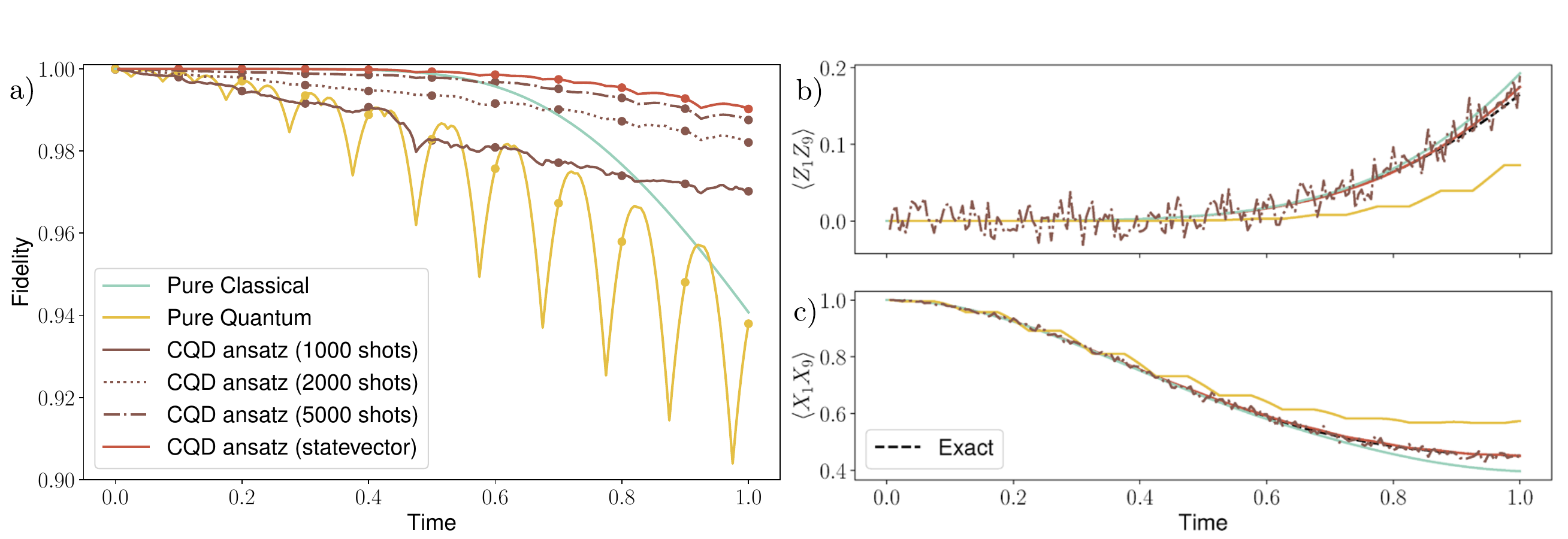}
    \caption{\textbf{Reducing the circuit depth} by simulating the TFIM setup displayed in~\Cref{fig:j1j2_explain} with the CQD ansatz, where the quantum circuit only includes nearest neighbor couplings and the classical correction is chosen as a Jastrow ansatz. The Hamiltonian parameters are $J_1=0.5$, $J_2= 0.1$ and $h=1$. The Trotter decomposition is chosen at second order with $\Delta t = 0.1$. (a) Fidelity of the quantum simulation neglecting next-nearest neighbor terms (yellow line), of the purely classical simulation (blue line) and of the CQD ansatz for different numbers of shots (red and brown lines). (b) $ZZ$ and (c) $XX$ correlation between the top left and bottom right spins in~\Cref{fig:j1j2_explain}. For visibility, we only include the simulation with 5000 shots.}
    \label{fig:j1j2_fid}
\end{figure*}

As an example, we consider a two-dimensional TFIM with nearest neighbor coupling $J_1$ and next-nearest neighbor coupling $J_2$ in a grid of $3\times3$ spins (see~\nobreak\Cref{fig:j1j2_explain}) 
\begin{align}\begin{split}
    \label{eq:j1j2_ham}
    H &= -h\sum_{i}X_i - J_1\sum_{\langle ij \rangle}Z_{i}Z_{j} - J_2 \sum_{\langle \langle  ij \rangle\rangle}Z_{i}Z_{j} \\
    &= \tilde{H}- J_2 \sum_{\langle \langle  ij \rangle\rangle}Z_{i}Z_{j}.
\end{split}\end{align}

We assume the nearest neighbor interaction is much stronger $J_1\gg J_2$ such that the part of the Hamiltonian that only includes the local field and nearest neighbor couplings $\tilde{H}$ is a good approximation of the total Hamiltonian $H$. In this case, we can employ simple classical models (such as a Jastrow ansatz~\cite{jastrow1955}) to correct the total dynamics with the hybrid CQD scheme. \Cref{fig:j1j2_explain} shows how the exclusion of non-local terms (yellow lines/gates) reduces the circuit depth significantly even given an all-to-all qubit connectivity. For a quantum device with nearest neighbor interactions, the benefit would be even larger as SWAP gates would be required to implement all the next-nearest neighbor interacting terms not included in $\tilde{H}$.

The initial state is chosen as $\ket{\psi_0} = \ket{+}^{\otimes 9}$ and the classical parameters are initialized zero. To implement the dynamics as a quantum circuit, the time evolution is Trotterized to second order with $\Delta t = 0.1$, where the CQD ansatz is additionally tasked with correcting the Trotter error as described in~\Cref{sec:apps_trott}. We use the Euler method with $\delta t = 0.005$ to integrate the equations of motion. 

\Cref{fig:j1j2_fid} shows the fidelities and two correlated observables of the simulation with the CQD ansatz compared to a purely quantum simulation (ignoring next-nearest neighbor terms), and a purely classical version. 
The hardware efficient circuit alone (yellow line) is not able to capture the full dynamics. This is to be expected given that the next-nearest neighbor terms are completely neglected in that case. Additionally, the effect of the Trotterization is apparent through an oscillating fidelity and piecewise continuous expectation values of the correlated observables $\langle Z_1 Z_9\rangle$ (panel (b)) and $\langle X_1X_9\rangle$ (panel (c)). On the other hand, the Jastrow ansatz optimized purely classically (blue line) performs better but also fails to provide a faithful simulation for longer times due to its limited expressivity.

Finally, the CQD ansatz combining the quantum circuit with the Jastrow ansatz achieves the highest fidelity. We plot the statevector simulation (red line) and compare it to shot based simulations (brown lines) with different numbers of shots per evaluated circuit. Note that while in these cases the ansatz is optimized with shot noise, the fidelity is always computed using the full state. As expected, the quality of the simulation improves as the number of shots increases. However, already for a modest number of 1000 shots, we observe an improvement in fidelity at later times over the classical Jastrow ansatz and the hardware efficient circuit alone. These results suggest that we are indeed able to correct the quantum circuit evolution for missing terms in the Hamiltonian and, consequently, reduce the circuit depth as well as avoid SWAP gates.

\subsection{Extending the system size}
\label{sec:apps_2subs}
In this section we discuss how the hybrid ansatz can be employed to expand the system size by including additional degrees of freedom in the classical model. This is particularly useful, if the system can be split into two subsystems where one subsystem is highly correlated and benefits from the simulation on a quantum computer whereas the second subsystem is only weakly correlated and can be simulated classically. The extended CQD ansatz takes as input configurations of both the highly correlated part $z_q$ and the weakly correlated bath $z_c$ and is defined as
\begin{align}
    \begin{split}
    \label{eq:ansatz_2subsystems}
    \psi_{\theta(t + \delta t)}&\left(t+ \delta t,z_c, z_q\right) \\
    &\qquad\qquad= \psi_{\theta(t)+\delta \theta}^c(z_c, z_q)\bra{z_q}e^{-i\delta t\tilde{H}}\ket{\psi^q(t)}\\
    &\qquad\qquad= \psi_{\theta(t)+\delta \theta}^c(z_c,z_q)\psi^q(t,z_q),
    \end{split}
\end{align}
where $\tilde{H}$ is a Hamiltonian that only acts on the highly correlated subsystem simulated on the quantum device. For a general Hamiltonian of the form $H = H_q + H_c + H_{qc}$ where $H_c$ ($H_q$) acts on the classically treated bath (quantum) degrees of freedom and $H_{qc}$ acts on both systems, a straightforward choice is $\tilde{H} = H_q$.

When computing expectation values of observables with the extended CQD ansatz, we can no longer sample the full configurations over the whole system on the quantum device. Instead, we sample only $z_q$ from the quantum circuit and use alternative, classical ways of sampling the bath configurations $z_c$. 
To this end, we redefine the classical contribution of the wave function as
\begin{equation}
    \label{eq:split_classical}
    \psi^c_\theta(z_c,z_q) = \psi^{c_1}_\theta(z_q)\psi^{c_2}_\theta(z_q,z_c),
\end{equation}
where the factor depending on $z_c$ is constrained to be normalized according to
\begin{equation}
\label{eq:normalisation_2sub}
    \sum_{z_c}|\psi^{c_2}_\theta(z_c, z_q)|^2 = 1, \quad \forall z_q.
\end{equation}
Using this ansatz, $z_c$ can be directly sampled from $|\psi^{c_2}_\theta(z_c, z_q)|^2$ given $z_q$. Further, to calculate the total normalization appearing in equations~\eqref{eq:unnorm_s}~-~\eqref{eq:unnorm_t}, no sampling of the classical configuration $z_c$ is required since
\begin{align}\begin{split}
    \label{eq:norm_2sub}
    \braket{\psi_\theta} &= \sum_{z_q,z_c}|\psi_\theta(z_q,z_c)|^2 \\
    &= \sum_{z_q}|\psi^q(z_q)|^2|\psi^{c_1}_\theta(z_q)|^2\sum_{z_c}|\psi^{c_2}_\theta(z_c,z_q)|^2 \\
    &=\sum_{z_q}|\psi^q(z_q)|^2|\psi^{c_1}_\theta(z_q)|^2.
\end{split}\end{align}
In~\Cref{app:2sub_terms}, we show how these properties allow us to compute all the quantities in the TDVP equations of motion in a similar fashion as discussed in~\Cref{sec:calc_forc_and_qgt}.

\begin{figure}
    \centering
    \includegraphics[width=\linewidth]{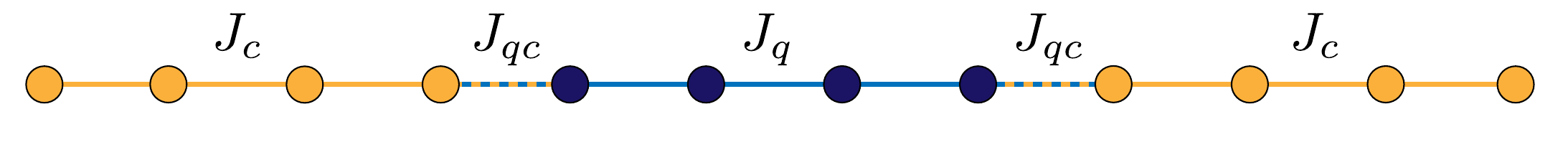}
    \caption{Spin chain in a TFIM split into four strongly correlated spins in the middle (blue), which are treated on a quantum device. The remaining (bath) spins (yellow) are weakly correlated and treated purely classically.}
    \label{fig:2sub_sketch}
\end{figure}

To test the extended CQD ansatz we simulate a TFIM on a line partitioned into two subsystems (see~\Cref{fig:2sub_sketch}). Four central spins, referred to as the \textit{quantum partition}, are connected with nearest neighbor coupling strength $J_q = 1$. Both ends of the chain are further connected to a bath chain of $n_c/2$ spins with an interaction strength $J_{qc} = 0.25$. The bath spins themselves are connected with a nearest neighbor coupling of $J_c = 0.1$ and can thus be regarded as only weakly correlated. The local fields are set to $h=1$ over the whole system. On the quantum device, we evolve the central spins according to $\tilde{H}$, which contains the terms of $H$ that only act on this partition (excluding the coupling between quantum and bath subsystems).

We choose the same Jastrow ansatz \eqref{eq:jastrow} as in the previous sections for the classical ansatz $\psi^{c_1}_\theta(z_q)$ acting only on the strongly correlated spins. The bath chains are represented by a mean-field ansatz $\psi^{c_2}_\theta$ defined as
\begin{equation}
    \label{eq:hybrid_mf}
    \psi^{c_2}_\theta(z_q,z_c) = \prod_{i=1}^{n_c} \frac{e^{\lambda_\theta(z_q)_iz_c^{(i)}}}{\sqrt{e^{2\Re{\lambda_\theta(z_q)_i}} + e^{-2\Re{\lambda_\theta(z_q)_i}}}},
\end{equation}
where $\lambda_\theta$ is a function parameterized by $\theta$ and taking a quantum configuration $z_q$ as input. As a consequence, the parameters of the mean-field ansatz are no longer fixed but depend on the configurations sampled from the quantum circuit $z_q$ through the function $\lambda_\theta$. This dependence serves as a backaction between the two subsystems and is essential to allow for correlations between them. We choose a neural network with one hidden layer with $2n_c$ neurons and $\tanh$ activation function to map $z_q \mapsto \lambda_\theta(z_q)$. During the time evolution, we hence have to update both the parameters in the Jastrow ansatz $\psi^{c_1}_\theta$ and the weights and biases of this neural network.

The mean-field ansatz allows for exact sampling of $z_c$ given $z_q$ by sampling each spin individually from the normalized distribution
\begin{equation}
    \label{eq:spin_sample_mf}
    p_i(z_c^{(i)}) = \frac{e^{2\Re{\lambda_\theta(z_q)_iz_c^{(i)}}}}{e^{2\Re{\lambda_\theta(z_q)_i}} + e^{-2\Re{\lambda_\theta(z_q)_i}}}.
\end{equation}

The initial state is once again chosen as $\ket{\psi_0} = \ket{+}^{\otimes (n_q + n_c)}$. We set the parameters of $\psi^{c_1}_\theta$ as well as the weights and biases of the final layer of the neural-network to zero to ensure that $\psi^c_{\theta(0)}(z_c,z_q) \equiv 1$ at time $t=0$. The remaining parameters are initialized randomly to avoid vanishing forces. We use second order Trotterization with $\Delta t = 0.1$ and integrate the equations of motion using the Euler method with $\delta t = 0.001$. 

\begin{figure*}[t!]
    \centering
    \includegraphics[width=0.8\linewidth]{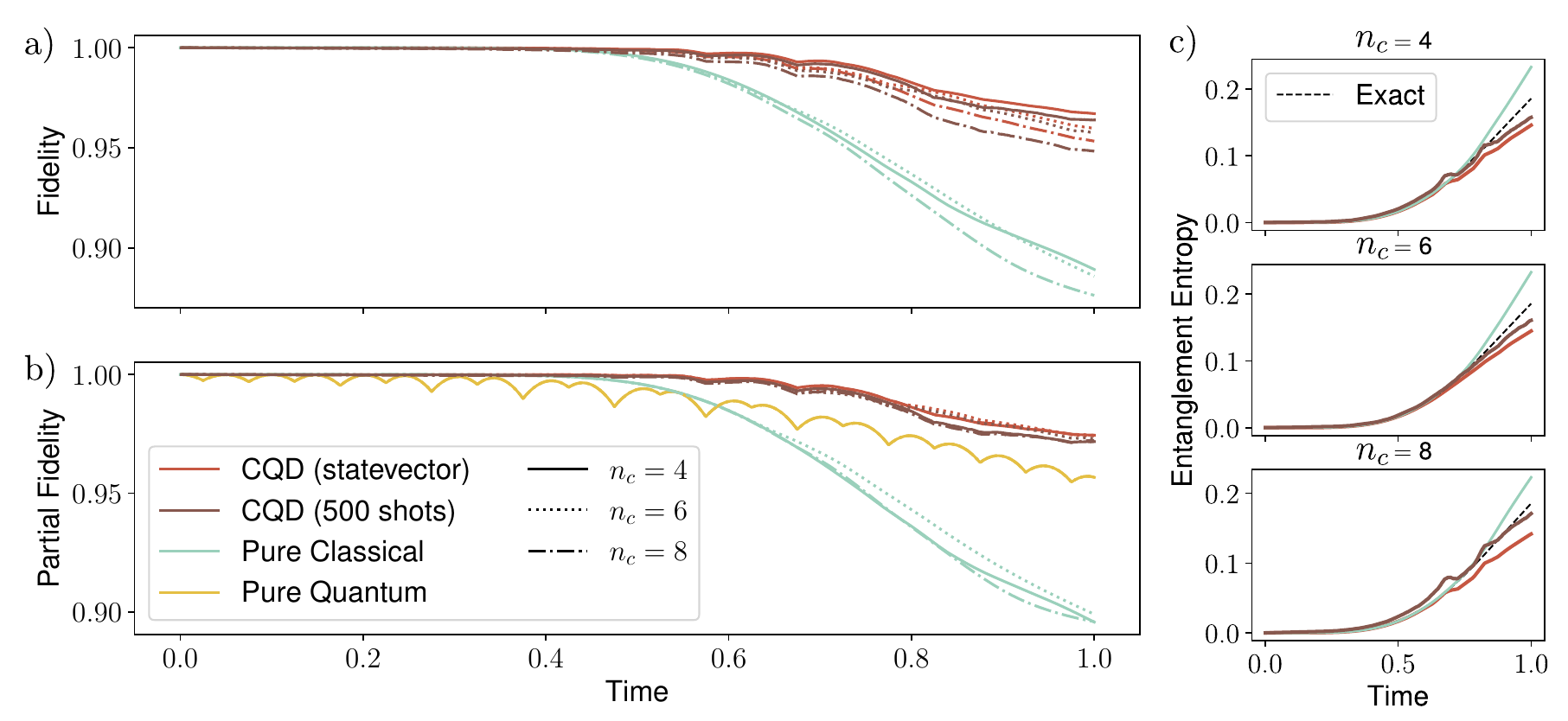}
    \caption{\textbf{Extending the system size} by simulating the TFIM on two subsystems as portrayed in~\Cref{fig:2sub_sketch} using a Jastrow ansatz on the quantum partition and a mean-field ansatz for the classical spins. We compare how the CQD ansatz performs compared to a purely classical simulation with the same classical ansatz for different number of bath spins $n_c$. (a) Global fidelity compared to the exact solution. (b) Fidelity of the reduced states, where the bath spins have been traced out. Here, we also include the fidelity of the Trotterized quantum circuit without the classical correction (yellow line). In (c), we plot the entanglement entropy between the two subsystems and compare it to the exact solution in black.}
    \label{fig:2sub_fids}
\end{figure*}

In~\Cref{fig:2sub_fids}~(a), we plot the fidelity of the CQD simulation with respect to the exact solution for different number of bath spins $n_c$ (red and brown lines). In addition, we include the fidelity obtained by optimizing the same classical ansatz $\psi^c_\theta(z_c, z_q)$ without the help of a quantum circuit using TDVP provided by the NetKet Python library~\cite{netket3:2022} (blue lines). The purely classical simulation achieves high fidelities up to $t \lesssim 0.5$, when the fidelity starts to significantly drop for all considered system sizes. In contrast, for the CQD ansatz the fidelity stays high for the whole evolution. These results again demonstrate that incorporating the quantum circuit into a simple classical ansatz increases its expressivity, allowing for more accurate simulations at longer times.

In addition to the statevector simulation, we perform a shot-based simulation with the CQD ansatz (brown lines). To that end, we sample 500 configurations $z_q$ from every circuit. Given $z_q$, we then sample 10 classical configurations $z_c$ from the mean-field ansatz. \Cref{fig:2sub_fids} shows that the shot-based optimization is able to reach a comparable fidelity to the statevector results indicating that a modest number of shots (and similarly a small number of classical samples) is sufficient for the system sizes considered here.

Note that we observe a slight decrease in the accuracy of the simulation for increasing bath sizes $n_c$ which, however, is to be expected as the classical bath partition is modeled using a mean-field ansatz and thus, the many-body fidelity naturally decreases with the bath size. Hence, to further study the system size dependency and simulation accuracy on the quantum partition alone, we compute the partial fidelity of the evolved quantum state (see~\Cref{fig:2sub_fids}~(b)), where we trace out the bath spins
\begin{equation}
    \label{eq:partial_fid}
    F_q(\rho,\sigma) = \text{Tr}\left[\sqrt{\sqrt{\rho_q}\sigma_q\sqrt{\rho_q}}\right]^2,
\end{equation}
where $\rho_q = \text{Tr}_c[\rho] = \text{Tr}_c[\ket{\psi}\bra{\psi}]$ is the reduced state. The partial fidelity of the purely classical simulation drops after $t \sim 0.5$ similar to the global fidelity. The purely quantum evolution that ignores the bath spins (yellow line) shows the usual oscillatory behavior due to the Trotterization with a partial fidelity that slowly decreases on average but already obtains superior results compared to the classical simulation. Finally, the CQD ansatz yields the highest partial fidelities throughout the time evolution. In contrast to the global fidelity, the partial fidelity reached by CQD remains unaffected by the number of bath spins included in the simulation, both for the statevector and shot based simulations. Moreover, the improved accuracy of the hybrid CQD ansatz compared to the purely quantum evolution demonstrates that the classical ansatz allows for information to propagate back from the bath into the quantum partition and thus is able to capture correlations between the subsystems.

To further study this correlation, in~\Cref{fig:2sub_fids}~(c), we plot the von Neumann entanglement entropy between the quantum and classical partition defined on the reduced density matrix $\rho_q$~\cite{nielsen_chuang_2010} 
\begin{equation}
    E[\rho_q] = -\text{Tr}\left[\rho_q\log(\rho_q)\right].
\end{equation}
We find that the backaction between the bath and quantum partitions introduced with the neural network indeed allows to recover the entanglement between the two subsystems, independent of the size of the bath.

Overall, these results demonstrate that the CQD ansatz enables an extension of the system size through a simple classical ansatz, which alone would lack the expressiveness required for accurately simulating the dynamics over longer timescales.

\section{Discussion \& Outlook}
\label{sec:conclusion}
In this work, we introduce a method that leverages classical simulation techniques to correct and extend Trotterized dynamics of quantum many-body systems on near-term quantum hardware. By exploiting the strengths of quantum computers — such as their ability to capture growing entanglement during time evolution — while relying on classical methods to mitigate their limitations, such as restricted connectivity and limited coherence time, we provide a new, near-term quantum-classical algorithm for the simulation of quantum dynamics.

The key principle of our method is that the quantum circuit is evolved with respect to an approximation of the full Hamiltonian. This approximation is then corrected by iteratively updating the parameters of a classical ansatz using TDVP. To obtain the correct equations of motion, we extend TDVP to the case where the ansatz contains an explicit time dependence. Crucially, the subroutine run on the quantum computer only consists of Trotterized time evolution and sampling configurations in easily preparable bases. Since the quantum circuit has no variational parameters, the forces and the quantum geometric tensor required for TDVP can be evaluated purely classically, given samples from the quantum device. This eliminates the need for computationally demanding tasks such as calculating complex overlaps (e.g., using the Hadamard test) or gradients (e.g., using the parameter-shift rule) on a quantum computer.

We demonstrate that incorporating a simple Jastrow ansatz in a coarsely Trotterized dynamics simulation of the TFIM significantly improves the quantum state fidelity. This shows that our CQD framework can be employed to correct Trotter errors often arising in digital quantum simulations with near-term hardware.
Furthermore, we show that the ansatz allows for the simulation of long-ranged Hamiltonians while keeping the quantum circuit hardware efficient. To that end, we simulate a two-dimensional TFIM with strong nearest neighbor and weaker next-nearest neighbor coupling strengths. Finally, the CQD ansatz allows to increase the system size beyond the available qubits of the quantum device. We demonstrate this by using a classical mean-field ansatz to incorporate additional spins that are treated purely classically. Each of these applications highlights the capability of the hybrid CQD ansatz in enhancing the accuracy of approximate dynamics simulation on quantum hardware. 

Moreover, we show that the CQD ansatz leverages the approximate simulation on the quantum device to increase the expressivity compared to a purely classical ansatz. This enables the use of simpler classical ansätze with fewer parameters to simulate highly entangled systems, where such ansätze would otherwise fail. Reducing the number of variational parameters is particularly beneficial, as the inversion of the quantum geometric tensor is prone to instabilities when employing complex ansätze with a large number of parameters~\cite{Hofmann22}.

In this article, we touch upon three applications of the method. However, these are by no means exhaustive. On the contrary, the proposed ansatz opens the door to any system, where the time evolution can only be implemented approximately on the quantum device. As long as the effective Hamiltonian acting on the quantum partition is known, our framework can be used to correct the dynamics with a sufficiently expressive classical ansatz. Furthermore, the ability of the CQD ansatz to extend system sizes—provided the additional degrees of freedom are efficiently classically simulable—suggests its applicability to a variety of physically interesting systems. Examples include molecular systems that can be partitioned into active and inactive orbitals~\cite{liu2019vqe,McArdle2020} or quantum impurity models~\cite{Kotliar_2006, Sun2016}. 
Finally, note that our derivation of TDVP with an explicit time dependence is general and can be employed in any dynamics simulation—classical or quantum—and with any parameterized ansatz that contains an explicit time dependence.

There are numerous potential extensions and improvements to the CQD ansatz. The framework itself is kept deliberately general, particularly with regard to the choice of classical ansatz. While we primarily discuss the Jastrow ansatz, alternative classical wave functions—such as tensor networks~\cite{SCHOLLWOCK201196} or neural network quantum states~\cite{carleo2017}—could also be considered. Additionally, a scaling analysis using near-term quantum hardware would be valuable in assessing the method's performance under realistic noise conditions. 

Currently, the main bottleneck of the CQD framework is the fact that quantum circuits have to be sampled frequently, as the TDVP time step has to be chosen small enough to allow for a stable and accurate numerical integration of the equations of motion. Hence, exploring alternative ways of sampling the quantum circuits is an interesting direction for future research.

\vspace{-3mm}
\section*{Code availability}
The code to run the CQD ansatz is available as a python package on github~\cite{cqd_code}. The repository contains the simulations performed in this article as examples and provides a general framework to simulate the dynamics of arbitrary Hamiltonians where the classical ansatz can easily be interchanged. It is built upon jax~\cite{jax2018github}, flax~\cite{flax2020github}, NetKet~\cite{netket3:2022} and Pennylane~\cite{pennylane}.
\vspace{-3mm}
\begin{acknowledgments}
 We thank Stefano Barison, Jannes Nys, Linda Mauron, Shao-Hen Chiew and Alessandro Sinibaldi for insightful discussions on hybrid methods and the time-dependent variational principle for classical ansätze.  This research was supported by the NCCR MARVEL, a National Centre of Competence in Research, funded by the Swiss National Science Foundation (grant number 205602).
 \end{acknowledgments}

\bibliography{notes.bib}

\onecolumngrid

\appendix

\section{Measuring non-diagonal observables and forces}
\label{app:expvals}
In~\Cref{sec:methods}, we show how all the quantities required for the time evolution with the CQD ansatz as well as the measurement of observables can be written in the form
\begin{equation}
    E(P, f) = \sum_{z,z'} \psi^q(t,z)^*\psi^q(t,z')\bra{z}{P}\ket{z'}f(t,z, z').
\end{equation}
For diagonal Pauli strings, recasting this sum as a sampling problem is straightforward (see discussion in~\Cref{sec:hybrid_ansatz} of the main text). For off-diagonal Pauli strings, we employ the strategy proposed by Zhang~et~al.~\cite{Zhang22VarQNHybrid}. First note that a Pauli string $P$ maps every computational basis state $z$ to another computational basis state $\tilde{z}$ up to a phase $S(\tilde{z}) \in \{\pm 1,\pm i\}$. We thus write $Pz = S(\tilde{z})\tilde{z}$, which allows us to eliminate the sum over $z'$ as
\begin{align}\begin{split}
     E(P, f) &= \sum_{z,z'} \psi^q(z)^*\psi^q(z')\bra{\tilde{z}}S(z)\ket{z'}f(z, z') \\
     &= \sum_z \psi^q(z)^*\psi^q(\tilde{z})S(z)f(z,\tilde{z}) = \sum_z \bra{\psi^q}\ket{z}\bra{\tilde{z}}\ket{\psi^q}S(z)f(z,\tilde{z}) \\
     &= \bra{\psi^q} \left[\sum_zS(z)f(z,\tilde{z})\ket{z}\bra{\tilde{z}} \right]\ket{\psi^q},
\end{split}\end{align}
where we dropped the time dependence for readability and used $S(\tilde{z})^* = S(z)$. We now reorder the qubits such that the Pauli acting on the first qubit $P_0$ is non-diagonal, i.e.~either $X$ or $Y$. As such, if $z_0 = 0$ we have $\tilde{z}_0 = 1$ and vice-versa which allows us to rewrite the sum by fixing $z_0=0$ as
\begin{equation}
     E(P, f) = \bra{\psi^q} \left[\sum_{z,\text{s.t.} z_0 = 0}S(z)f(z,\tilde{z})\ket{z}\bra{\tilde{z}} + S(\tilde{z})f(\tilde{z},z)\ket{\tilde{z}}\bra{z} \right]\ket{\psi^q}.
\end{equation}

Each term in this sum can be written as
\begin{align}\begin{split}
f(z,\tilde{z})S(z)\ket{z}\bra{\tilde{z}} &+ f(\tilde{z},z)S(\tilde{z})\ket{\tilde{z}}\bra{z}  \\
&=\frac{1}{2}\left( f(z,\tilde{z}) + f(\tilde{z},z)\right)\left(S(z)\ket{z}\bra{\tilde{z}} + S(\tilde{z})\ket{\tilde{z}}\bra{z}\right)\\
&\quad+ \frac{1}{2}\left( f(z,\tilde{z}) - f(\tilde{z},z)\right)\left(S(z)\ket{z}\bra{\tilde{z}} - S(\tilde{z})\ket{\tilde{z}}\bra{z}\right) \\
&=\frac{1}{2}\left( f(z,\tilde{z}) + f(\tilde{z},z)\right)\\
&\quad\quad\quad\left(\ket{+,z_{1:n-1}}\bra{+,z_{1:n-1}} - \ket{-,z_{1:n-1}}\bra{-,z_{1:n-1}}\right) \\
&\quad+ \frac{i}{2}\left( f(z,\tilde{z}) - f(\tilde{z},z)\right)\\
&\quad\quad\quad\left(\ket{+i,z_{1:n-1}} \bra{+i,z_{1:n-1}} - \ket{-i,z_{1:n-1}} \bra{-i,z_{1:n-1}}\right),
\end{split}\end{align}
where
\begin{align}\begin{split}
    \ket{\pm, z_{1:n-1}} &= \frac{1}{\sqrt{2}}\left(S(z)\ket{0, z_{1:n-1}} \pm \ket{1,\tilde{z}_{1:n-1}}\right), \\
    \ket{\pm i, z_{1:n-1}} &= \frac{1}{\sqrt{2}}\left(S(z)\ket{0, z_{1:n-1}} \pm i \ket{1,\tilde{z}_{1:n-1}}\right).
\end{split}\end{align}
Note that these states can be easily prepared using controlled Pauli operations. In the case that $P_0 = X$ and hence $S(z_0) = 1$, we find
\begin{align}\begin{split}\label{eq:circ_x}
    \ket{+, z_{1:n-1}} &= \left(\ket{0}\bra{0}\otimes \mathbb{I}_{n-1} + \ket{1}\bra{1}\otimes P_{1:n-1}\right)\left(H \otimes \mathbb{I}_{n-1}\right)\left(\ket{0}\otimes\ket{z_{1:n-1}}\right) \\
    \ket{-, z_{1:n-1}} &= \left(\ket{0}\bra{0}\otimes \mathbb{I}_{n-1} + \ket{1}\bra{1}\otimes P_{1:n-1}\right)\left(H \otimes \mathbb{I}_{n-1}\right)\left(\ket{1}\otimes\ket{z_{1:n-1}}\right) \\
    \ket{+ i, z_{1:n-1}} &= \left(\ket{0}\bra{0}\otimes \mathbb{I}_{n-1} + \ket{1}\bra{1}\otimes P_{1:n-1}\right)\left(SH \otimes \mathbb{I}_{n-1}\right)\left(\ket{0}\otimes\ket{z_{1:n-1}}\right)\\
    \ket{- i, z_{1:n-1}} &= \left(\ket{0}\bra{0}\otimes \mathbb{I}_{n-1} + \ket{1}\bra{1}\otimes P_{1:n-1}\right)\left(SH \otimes \mathbb{I}_{n-1}\right)\left(\ket{1}\otimes\ket{z_{1:n-1}}\right),
\end{split}\end{align}
where $H$ is the Hadamard and $S$ the phase gate. For $P_0 = Y$ with $S(z_0) = i$, we need an extra $-S$ gate, resulting in
\begin{align}\begin{split}\label{eq:circ_y}
    \ket{+, z_{1:n-1}} &= \left(\ket{0}\bra{0}\otimes \mathbb{I}_{n-1} + \ket{1}\bra{1}\otimes P_{1:n-1}\right)\left((-S)H \otimes \mathbb{I}_{n-1}\right)\left(\ket{0}\otimes\ket{z_{1:n-1}}\right) \\
    \ket{-, z_{1:n-1}} &= \left(\ket{0}\bra{0}\otimes \mathbb{I}_{n-1} + \ket{1}\bra{1}\otimes P_{1:n-1}\right)\left((-S)H \otimes \mathbb{I}_{n-1}\right)\left(\ket{1}\otimes\ket{z_{1:n-1}}\right) \\
    \ket{+ i, z_{1:n-1}} &= \left(\ket{0}\bra{0}\otimes \mathbb{I}_{n-1} + \ket{1}\bra{1}\otimes P_{1:n-1}\right)\left(H \otimes \mathbb{I}_{n-1}\right)\left(\ket{0}\otimes\ket{z_{1:n-1}}\right)\\
    \ket{- i, z_{1:n-1}} &= \left(\ket{0}\bra{0}\otimes \mathbb{I}_{n-1} + \ket{1}\bra{1}\otimes P_{1:n-1}\right)\left(H \otimes \mathbb{I}_{n-1}\right)\left(\ket{1}\otimes\ket{z_{1:n-1}}\right).
\end{split}\end{align}

In summary, we are able to rewrite
\begin{align}\begin{split}
    E(P,f) &= \frac{1}{2}\sum_{z_{1:n-1}}|\bra{\psi^q}\ket{+,z_{1:n-1}}|^2\left(f(z,\tilde{z}) + f(\tilde{z},z)\right) \\
    &\quad-\frac{1}{2}\sum_{z_{1:n-1}}|\bra{\psi^q}\ket{-,z_{1:n-1}}|^2\left(f(z,\tilde{z}) + f(\tilde{z},z)\right) \\
    &\quad+\frac{i}{2}\sum_{z_{1:n-1}}|\bra{\psi^q}\ket{+i,z_{1:n-1}}|^2\left(f(z,\tilde{z}) + f(\tilde{z},z)\right)  \\
    &\quad-\frac{i}{2}\sum_{z_{1:n-1}}|\bra{\psi^q}\ket{-i,z_{1:n-1}}|^2\left(f(z,\tilde{z}) + f(\tilde{z},z)\right)
\end{split}\end{align}
which allows us to evaluate the expectation value by sampling $z$ from the time-evolved circuit in the basis corresponding to either~\Cref{eq:circ_x} or~\Cref{eq:circ_y}, and evaluating the classical function $f(z,\tilde{z})$ accordingly on those samples.

\section{Derivation of~\Cref{tab:pauli}}
\label{app:der_table}
In this appendix, we provide the derivations for~\Cref{tab:pauli} of the main text that describes how all the terms appearing in the equations of motion~\eqref{eq:eom_norm} can be written in the form of~\eqref{eq:pauli_expect}.

We start with the norm 
 \begin{equation}
     \braket{\psi} = \sum_z |\psi^q(t,z)|^2 |\psi^c_{\theta}(z)|^2,
 \end{equation}
 which corresponds to~\eqref{eq:pauli_expect} for ${P} = {I}$ and $f(t,z,z') = |\psi^c_{\theta}(z)|^2$. Moving on to the quantum geometric tensor, we find
 \begin{equation}
     \bra{\partial_l\psi}\ket{\partial_k\psi} = \sum_z |\psi^q(t,z)|^2 (\partial_l\psi^c_{\theta}(z))^*\partial_k\psi^c_{\theta}(z),
 \end{equation}
 which again corresponds to the identity Pauli with $f(t,z,z') = (\partial_l\psi^c_{\theta}(z))^*\partial_k\psi^c_{\theta}(z)$. Similarly, the results for $\bra{\partial_k\psi}\ket{\psi}$ and $\bra{\psi}\ket{\partial_k\psi}$ can be found.

 Moving on to the forces, we now have to decompose the Hamiltonian into Pauli strings $H = \sum_i w_i{P}_i$ to find
 \begin{equation}
     \bra{\partial_k\psi}H\ket{\psi} = \sum_iw_i\bra{\partial_k\psi}{P_i}\ket{\psi} =  \sum_iw_i\sum_{z,z'} \psi^q(t,z)^*\psi^q(t,z')\bra{z}{P_i}\ket{z'}(\partial_k\psi^c_\theta(z))^*\psi^c_\theta(z'),
 \end{equation}
corresponding to~\eqref{eq:pauli_expect}, where the circuit needs to be sampled for every Pauli string in the decomposition of $H$ and the function to be evaluated is $f(t,z,z') = (\partial_k\psi^c_\theta(z))^*\psi^c_\theta(z')$. Similarly, the results for $\bra{\psi}H\ket{\psi}$ can be found. 

The derivation for $\bra{\psi}\ket{\partial_t\psi}$ is given in the main text and the results for $\bra{\partial_k\psi}\ket{\partial_t\psi}$ are found equivalently.

\subsection{Derivation for the case of two subsystems}
\label{app:2sub_terms}
In this section, we provide the same calculations as above for the case where additional degrees of freedom are added purely classically. As in the main text, we assume the global ansatz is given as 
\begin{equation}
   \psi_{\theta(t + \delta t)}\left(t+ \delta t,z_c, z_q\right) =  \psi^q(t,z_q)\psi_{\theta(t)+\delta \theta}^c(z_c,z_q) = \psi^q(t,z_q)\psi_{\theta(t)+\delta \theta}^{c_1}(z_q)\psi_{\theta(t)+\delta \theta}^{c_2}(z_c,z_q)
\end{equation}
with
\begin{equation}
    1 = \sum_{z_c}|\psi^{c_2}_\theta(z_c,z_q)|^2 \quad \forall z_q.
\end{equation}

In addition to the challenge of sampling $z_c$ described in the main text, we have to consider Pauli strings that act on either or both partitions. We write ${P} = {P}^c \otimes {P}^q$ for Pauli strings appearing in the decomposition of $H$. Note that, in contrast, $\tilde{H}$ only acts on the quantum partition. We use the notation ${P}\ket{z} = S(\tilde{z})\ket{\tilde{z}}$ introduced in~\Cref{app:expvals} to denote the connected elements of a Pauli string ${P}$. We further assume that the classical functions are given in the form
\begin{equation}
    \psi^{c_1}_\theta(z_q) = \exp(\varphi^{c_1}_\theta(z_q)), \quad  \psi^{c_2}_\theta(z_c, z_q) = \exp(\varphi^{c_2}_\theta(z_c, z_q)),
\end{equation}
such that the derivatives can be written as
\begin{equation}
    \partial_k\psi^c_\theta(z_c, z_q) = \partial_k\exp\left(\varphi^{c_1}_\theta(z_q) + \varphi^{c_2}_\theta(z_c, z_q)\right) = \psi^c_\theta(z_c, z_q)\partial_k\left(\varphi^{c_1}_\theta(z_q) + \varphi^{c_2}_\theta(z_c, z_q)\right) = \psi^c_\theta(z_c, z_q)\partial_k\varphi^{c}_\theta(z_c, z_q),
\end{equation}
where we defined $\varphi^{c}_\theta(z_c, z_q) =\varphi^{c_1}_\theta(z_q) + \varphi^{c_2}_\theta(z_c, z_q)$.

With this notation, we can now write the force for $H = \sum_iw_i{P}_i$  as
\begin{align}\begin{split}
    \bra{\partial_k\psi}H\ket{\psi} &= \sum_iw_i\sum_{z_q, z_q'}\psi^q(z_q)^*\psi^q(z_q')\bra{z_q}{P}_i^q\ket{z_q'}\sum_{z_c, z_c'}\left(\partial_k\psi^c_\theta(z_c, z_q)\right)^*\psi^c_\theta(z_c', z_q')\bra{z_c}{P}^c_i\ket{z_c'} \\
    &= \sum_iw_i\sum_{z_q, z_q'}\psi^q(z_q)^*\psi^q(z_q')\bra{z_q}{P}_i^q\ket{z_q'}f_i(t,z_q,z_q'),
 \end{split}\end{align}
where the classical functions to be evaluated can be written as
\begin{align}\begin{split}
    f_i(t, z_q, z_q') &= \psi^{c_1}_\theta(z_q)^*\psi^{c_1}_\theta(z_q')\sum_{z_c, z_c'}\psi^{c_2}_\theta(z_c, z_q)^*\psi^{c_2}_\theta(z_c', z_q')\left(\partial_k\varphi^{c}_\theta(z_c, z_q)\right)^*\bra{z_c}{P}^c_i\ket{z_c'}\\
    &=\psi^{c_1}_\theta(z_q)^*\psi^{c_1}_\theta(z_q')\sum_{z_c}\psi^{c_2}_\theta(z_c, z_q)^*\psi^{c_2}_\theta(\tilde{z}_c, z_q')\left(\partial_k\varphi^{c}_\theta(z_c, z_q)\right)^*S_i(z_c) \\
    &=\psi^{c_1}_\theta(z_q)^*\psi^{c_1}_\theta(z_q')\sum_{z_c}|\psi^{c_2}_\theta(z_c, z_q)|^2\frac{\psi^{c_2}_\theta(\tilde{z}_c, z_q')}{\psi^{c_2}_\theta(z_c, z_q)}\left(\partial_k\varphi^{c}_\theta(z_c, z_q)\right)^*S_i(z_c) \\
    &=\psi^{c_1}_\theta(z_q)^*\psi^{c_1}_\theta(z_q')\mathbb{E}_{z_c \sim |\psi^{c_2}_\theta(z_c, z_q)|^2}\left[\frac{\psi^{c_2}_\theta(\tilde{z}_c, z_q')}{\psi^{c_2}_\theta(z_c, z_q)}\left(\partial_k\varphi^{c}_\theta(z_c, z_q)\right)^*S_i(z_c)\right].
\end{split}\end{align}
For each sampled $z_q$, we thus have to estimate an expectation value by sampling $z_c$ from $|\psi^{c_2}_\theta(z_c, z_q)|^2$. The remaining terms can be found similarly and are listed in~\Cref{tab:pauli_2sub}.
\begin{table}[h]
    \centering
    \renewcommand{\arraystretch}{1.5} 
    \begin{tabular}{|c|c|c|}
        \hline
        \textbf{Expression} & \textbf{Pauli Strings} & \boldmath$f(t, z_q, z_q')$ \\
        \hline
        $\braket{\psi}$ & Identity & $|\psi^{c_1}_\theta(z_q)|^2$ \\
        \hline
        $\bra{\partial_l\psi}\ket{\partial_k\psi}$ & Identity & $|\psi^{c_1}_\theta(z_q)|^2\mathbb{E}_{z_c \sim z_q}\left[(\partial_l\varphi^c_\theta(z_c, z_q))^*\partial_k\varphi^c_\theta(z_c, z_q)\right]$ \\
        \hline
        $\bra{\partial_k\psi}\ket{\psi}$ & Identity & $|\psi^{c_1}_\theta(z_q)|^2\mathbb{E}_{z_c \sim z_q}\left[(\partial_k\varphi^c_\theta(z_c, z_q))^*\right]$ \\
        \hline
        $\bra{\psi}H\ket{\psi}$ & ${P}^c_i \otimes {P}^q_i \in \mathcal{P}(H)$  & $\psi^{c_1}_\theta(z_q)^*\psi^{c_1}_\theta(z_q')\mathbb{E}_{z_c \sim z_q}\left[\frac{\psi^{c_2}_\theta(\tilde{z}_c, z_q')}{\psi^{c_2}_\theta(z_c, z_q)}S_i(z_c)\right]$ \\
        \hline
        $\bra{\partial_k\psi}H\ket{\psi}$ & ${P}^c_i \otimes {P}^q_i \in \mathcal{P}(H)$  & $\psi^{c_1}_\theta(z_q)^*\psi^{c_1}_\theta(z_q')\mathbb{E}_{z_c \sim z_q}\left[\frac{\psi^{c_2}_\theta(\tilde{z}_c, z_q')}{\psi^{c_2}_\theta(z_c, z_q)}\left(\partial_k\varphi^{c}_\theta(z_c, z_q)\right)^*S_i(z_c)\right]$ \\
        \hline
        $\braket{\psi}{\partial_t\psi}$ & $P_i^q \in \mathcal{P}(\tilde{H})$  & $|\psi^{c_1}_\theta(z)_q|^2$ \\
        \hline
        $\braket{\partial_k\psi}{\partial_t\psi}$ & $P_i^q \in \mathcal{P}(\tilde{H})$  & $|\psi^{c_1}_\theta(z_q)|^2\mathbb{E}_{z_c \sim z_q}\left[(\partial_k\varphi^c_\theta(z_c, z_q))^*\right]$ \\
        \hline
    \end{tabular}
\caption{All the expressions appearing in the equations of motion can be computed using the form provided in~\eqref{eq:pauli_expect} for the case of two subsystems described in~\Cref{sec:apps_2subs}. Here, we list the Pauli strings according to which the circuit has to be sampled and the classical functions that have to be evaluated to compute each expression. We write $\mathcal{P}(H)$ to denote the set of Pauli strings appearing in the decomposition of $H$. We use the shorthand $z_c \sim z_q$ for $z_c$ sampled from $|\psi^{c_2}_\theta(z_c, z_q)|^2$ given $z_q$.}
    \label{tab:pauli_2sub}
\end{table}

\section{TDVP for unnormalized wave functions}
\label{app:unnormalized}
The derivation of the TDVP equations in the main text assumes that the ansatz remains normalized over time. To generalize this to unnormalized classical wave functions and to address the problem of the global phase being ignored in McLachlan's principle, we add two additional virtual parameters to the ansatz which will modify the equations of motion. We add the parameter $\theta_0^R$ to encode the normalization and $\theta_0^I$ to fix the global phase. The total ansatz now takes the form
\begin{equation}
        \label{eq:ansatz_norm}
    \psi_{\theta(t)}(t,z) = e^{\theta_0^R(t) + i\theta_0^I(t)}\psi_{\theta(t)}^c(z)\psi^q(t,z),
\end{equation}
where the classical ansatz $\psi_{\theta(t)}^c$ still only depends on the original parameters $\theta$.
To derive the new equations of motion for $\theta_{k>1}$, we first observe that
\begin{equation}
    |\partial_{\theta_0^R}\psi\rangle = \sum_z\partial_{\theta_0^R}\psi_{\theta(t)}(t,z)\ket{z} = \sum_z\psi_{\theta(t)}(t,z)\bra{z} = \ket{\psi}
\end{equation}
and similarly
\begin{equation}
    |\partial_{\theta_0^I}\psi\rangle = i\ket{\psi}.
\end{equation}
The equations of motion in~\eqref{eq:eom} for the virtual parameters $k=0$ are thus given by
\begin{align}\begin{split}
    \dot{\theta}_0^R\braket{\psi} &+ \sum_{l>0}\dot{\theta}_l \Re{\bra{\partial_l\psi}\ket{\psi}} = -\Re{\bra{\psi}\ket{\partial_t\psi}} \\
    \dot{\theta}_0^I\braket{\psi} &- \sum_{l>0}\dot{\theta}_l \Im{\bra{\partial_l\psi}\ket{\psi}} = -\bra{\psi}H\ket{\psi} - \Im{\bra{\psi}\ket{\partial_t\psi}}.
\end{split}\end{align}
We can now plug these results into the equation of motions for $k>0$ to obtain for the left hand side
\begin{align}\begin{split}
    \sum_l\dot{\theta}_l \Re{\bra{\partial_l\psi}\ket{\partial_k\psi}} &= \dot{\theta}_0^R\Re{\bra{\psi}\ket{\partial_k\psi}} + \dot{\theta}_0^I\Im{\bra{\psi}\ket{\partial_k\psi}} + \sum_{l>0}\dot{\theta}_l \Re{\bra{\partial_l\psi}\ket{\partial_k\psi}} \\
    &=-\sum_{l>0}\dot{\theta}_l \frac{\Re{\bra{\partial_l\psi}\ket{\psi}}\Re{\bra{\psi}\ket{\partial_k\psi}}}{\braket{\psi}} - \frac{\Re{\bra{\psi}\ket{\partial_t\psi}}\Re{\bra{\psi}\ket{\partial_k\psi}}}{\braket{\psi}} \\
    &\quad+\sum_{l>0}\dot{\theta}_l \frac{\Im{\bra{\partial_l\psi}\ket{\psi}}\Im{\bra{\psi}\ket{\partial_k\psi}}}{\braket{\psi}} - \frac{\bra{\psi}H\ket{\psi}\Im{\bra{\psi}\ket{\partial_k\psi}}}{\braket{\psi}} \\
    &\quad - \frac{\Im{\bra{\psi}\ket{\partial_t\psi}}\Im{\bra{\psi}\ket{\partial_k\psi}}}{\braket{\psi}} + \sum_{l>0}\dot{\theta}_l \Re{\bra{\partial_l\psi}\ket{\partial_k\psi}} \\
    &= \sum_{l>0}\dot{\theta}_l \left(\Re{\bra{\partial_l\psi}\ket{\partial_k\psi}} - \frac{\Re{\bra{\partial_l\psi}\ket{\psi}\bra{\psi}\ket{\partial_k\psi}}}{\braket{\psi}}\right) \\
    &\quad - \frac{\Re{\bra{\partial_k\psi}\ket{\psi}\bra{\psi}\ket{\partial_t\psi}}}{\braket{\psi}} + \frac{\Im{\bra{\partial_k\psi}\ket{\psi}\bra{\psi}H\ket{\psi}}}{\braket{\psi}}
\end{split}\end{align}
Setting this equal to the right hand side of~\eqref{eq:eom}, dividing both sides by $\braket{\psi}$ and rearranging yields the equations of motion for an unnormalized ansatz
\begin{equation}
    \sum_{l>0}S_{kl}\dot{\theta}_l = C_k - T_k,
\end{equation}
where 
\begin{equation}
    S_{kl} = \Re{\frac{\bra{\partial_l\psi}\ket{\partial_k\psi}}{\braket{\psi}} - \frac{\bra{\partial_l\psi}\ket{\psi}\bra{\psi}\ket{\partial_k\psi}}{\braket{\psi}^2}}
\end{equation}
is the geometric tensor,
\begin{equation}
C_k = \Im{\frac{\bra{\partial_k\psi}H\ket{\psi}}{\braket{\psi}} - \frac{\bra{\partial_k\psi}\ket{\psi}\bra{\psi}H\ket{\psi}}{\braket{\psi}^2}}
\end{equation}
are the usual forces and
\begin{equation}
T_k = \Re{\frac{\bra{\partial_k\psi}\ket{\partial_t\psi}}{\braket{\psi}} - \frac{\bra{\partial_k\psi}\ket{\psi}\bra{\psi}\ket{\partial_t\psi}}{\braket{\psi}^2}}
\end{equation}
are additional forces arising from the explicit time dependence of the ansatz.

\end{document}